# Intrinsic and Extrinsic Performance Limits of Graphene Devices on SiO$_2$

J.H. Chen[1,3], C. Jang[2,3], S. Xiao[2,3], M. Ishigami[3], and M. S. Fuhrer[1,2,3]

*[1]Materials Research Science and Engineering Center, [2]Center for Nanophysics and Advanced Materials, [3]Department of Physics, University of Maryland, College Park, MD 20742 USA*

**The linear dispersion relation in graphene[1,2] gives rise to a surprising prediction: the resistivity due to isotropic scatterers (e.g. white-noise disorder[3] or phonons[4-8]) is independent of carrier density *n*. Here we show that acoustic phonon scattering[4-6] is indeed independent of *n*, and places an *intrinsic* limit on the resistivity in graphene of only 30 Ω at room temperature (RT). At a technologically-relevant carrier density of $10^{12}$ cm$^{-2}$, the mean free path for electron-acoustic phonon scattering is >2 microns, and the intrinsic mobility limit is $2 \times 10^5$ cm$^2$/Vs, exceeding the highest known inorganic semiconductor (InSb, ~$7.7 \times 10^4$ cm$^2$/Vs[9]) and semiconducting carbon nanotubes (~$1 \times 10^5$ cm$^2$/Vs[10]). We also show that *extrinsic* scattering by surface phonons of the SiO$_2$ substrate[11,12] adds a strong temperature dependent resistivity above ~200 K[8], limiting the RT mobility to ~$4 \times 10^4$ cm$^2$/Vs, pointing out the importance of substrate choice for graphene devices[13].**

The nature of electron-phonon scattering in graphene has been determined by measuring the four-probe resistivity $\rho(V_g,T)$ of graphene field-effect devices on SiO$_2$/Si[2, 14] vs. temperature $T$ from 16 K – 485 K, and gate voltage $V_g$ applied to the Si substrate (see Methods). Measurements are performed in ultra-high vacuum (UHV) on cleaned samples to minimize temperature-dependent effects due to molecular adsorption/desorption[14, 15].

The dependence of resistivity on carrier density is investigated by using the gate voltage to tune the carrier density $n = c_g V_g/e$, where $c_g = 1.15 \times 10^{-8}$ F/cm$^2$ is the gate capacitance, and $e$ the elementary charge. Figure 1a and 1b show $\rho(V_g,T)$ for two samples at seven different gate voltages plotted on a linear scale. The $\rho(V_g,T)$ curves are linear in temperature at low $T$ with a slope of $(4.0 \pm 0.5) \times 10^{-6}$ $h/e^2 K$ as indicated by the short-dashed lines. The slope is independent of carrier density, and is the same for both samples.

Acoustic phonon scattering is expected[4-6, 16] to give rise to a linear resistivity independent of carrier density

$$\rho(V_g,T) = \rho_0(V_g) + \rho_A(T); \qquad \rho_A(T) = \left(\frac{h}{e^2}\right) \frac{\pi^2 D_A^2 k_B T}{2h^2 \rho_s v_s^2 v_F^2}, \qquad (1)$$

where $\rho_0(V_g)$ is the residual resistivity at low temperature, $\rho_A(T)$ is the resistivity due to acoustic phonon scattering, $k_B$ is the Boltzmann constant, $\rho_s = 6.5 \times 10^{-7}$ kg/m$^2$ is the 2D mass density of graphene, $v_F = 10^6$ m/s is the Fermi velocity, $v_s$ is the sound velocity, and $D_A$ the acoustic deformation potential. For LA phonons, $v_s = 2.1 \times 10^4$ m/s and our experimentally determined slope gives $D_A = 17 \pm 1$ eV, in good agreement with theoretical[6, 16-19] and experimental[20-22] expectations. (At very low temperature $T \ll T_{BG} \approx (v_s/v_F)T_F$, where $T_F$ is the Fermi temperature, a crossover to $\rho_A(T) \propto T^4$ is expected[6]; $T_{BG}$

≈ (8 K)$V_g^{1/2}$ where $V_g$ is measured in V, so this condition is reasonably satisfied in our experiments.)

In contrast to the low-$T$ behavior, the resistivity at higher $T$ is highly non-linear in $T$, and becomes significantly dependent on $V_g$, increasing for decreasing $V_g$. Morozov, et al.[8] noted the non-linear dependence on $T$ but were unable to separate the low-$T$ LA phonon contribution from the high-$T$ contribution,nor to identify the specific dependences on $T$ or $V_g$ for each contribution. The strong (activated) temperature dependence suggests scattering by a high-energy phonon mode or modes. We find that the data can be fitted by adding an extra term $\rho_B(V_g,T)$ representing the activated contribution to the resistivity:

$$\rho(V_g,T) = \rho_0(V_g) + \rho_A(T) + \rho_B(V_g,T); \quad \rho_B(V_g,T) = B_1 V_g^{-\alpha_1}\left(\frac{1}{e^{(59 meV)/k_BT}-1} + \frac{6.5}{e^{(155 meV)/k_BT}-1}\right)$$
, (2a)

or

$$\rho(V_g,T) = \rho_0(V_g) + \rho_A(T) + \rho_B(V_g,T); \quad \rho_B(V_g,T) = B_2 V_g^{-\alpha_2}\left(\frac{1}{e^{E_0/k_BT}-1}\right), \quad (2b)$$

For Eqn. 2a, the particular form of the expression in parenthesis in $\rho_B(V_g,T)$ is chosen to match surface phonons in $SiO_2$[12]; however, a single Bose-Einstein (BE) distribution as shown in Eqn. 2b can also give a reasonable fit. Figures 1c and 1d show a global fit to Eqn. 2a (solid lines) and to Eqn. 2b (short-dashed lines) to the data for two samples. The global best fit parameters are $B_1$ = 0.607 $(h/e^2)V^{\alpha_1}$ and $\alpha_1$ = 1.04 for Eqn. 2a, and $B_2$ = 3.26 $(h/e^2)V^{\alpha_2}$, $\alpha_2$ = 1.02, and $E_0$ = 104 meV for Eqn. 2b.

We now discuss the possible origins of the activated resistivity term $\rho_B(V_g,T)$. Scattering in graphene requires a phonon wavevector $q \approx 0$ (intravalley scattering) or $q \approx$ K (intervalley scattering). The next lowest-energy modes after the $q \approx 0$ acoustic modes are the zone boundary ZA phonon ($q = K$) at $\hbar\omega \approx 70$ meV and the optical ZO mode ($q = 0$) at $\hbar\omega \approx 110$ meV[23]. The optical ZO mode is consistent with the observed temperature dependence as per the fit to Eqn. 2b, however both modes are out-of-plane vibrations, which are not expected to couple strongly to the electrons[17-20]; for example scattering by these modes is not observed in carbon nanotubes, while scattering by the longitudinal zone-boundary phonon with $\hbar\omega \approx 160$ meV is extremely strong[24] (but the data are poorly fit to a BE distribution with $\hbar\omega \approx 160$ meV). The strong carrier density dependence $\rho_B(V_g,T) \propto V_g^{-1.04}$ is also inconsistent with graphene optical phonon scattering, which should depend very weakly on carrier density[6]. Breaking of the inversion symmetry of the graphene sheet by the substrate induces an additional perturbation potential for the out-of-plane phonon modes, but reasonable estimates of the size of this perturbation are too small to account for the observed $\rho_B(V_g,T)$. Thus we reject optical phonon modes of graphene as the source of $\rho_B(V_g,T)$.

Another possible origin of $\rho_B(V_g,T)$ is remote interfacial phonon (RIP) scattering[11] by the polar optical phonons of the $SiO_2$ substrate. This has been recently discussed theoretically in the context of graphene by Fratini and Guinea[12]. The two surface optical phonon modes in $SiO_2$ have $\hbar\omega \approx 59$ meV and 155 meV, with a ratio of coupling to the electrons of 1:6.5; we used these parameters as inputs to Eqn. 2a above, and the fit shows that they reasonably describe the temperature dependence of $\rho_B(V_g,T)$ (see Figs. 1c,d). The magnitude of the RIP scattering resistivity predicted by Fratini and Guinea[12] is on order a few $10^{-3}$ $h/e^2$ at 300 K, also in agreement with the observed magnitude. RIP results in a long-ranged potential, which gives rise to a density-dependent resistivity in graphene, similar to charged impurity scattering. Specifically, in the simplest case, the electron-phonon matrix $|H_{kk'}|^2$ element is proportional to $q^{-1}$ where $q$ is the scattering wavevector, and the resistivity is proportional to $k_F^{-1} \propto V_g^{-1/2}$. However, finite-$q$ corrections to $|H_{kk'}|^2$ lead to a stronger dependence of $\rho_B(V_g,T)$ on $V_g$[12], so the observed $\rho_B(V_g,T) \propto V_g^{-1.04}$ is also reasonable. RIP scattering[11, 12] by the polar optical phonons of the SiO$_2$ substrate therefore naturally explains the magnitude, temperature dependence, and charge carrier density dependence of $\rho_B(V_g,T)$, hence we consider RIP scattering[11] to be the most likely origin of $\rho_B(V_g,T)$.

The contributions of the acoustic phonons and remote interfacial phonons can be used to determine the room-temperature intrinsic limits to the resistivity and mobility in graphene, and extrinsic limits for graphene on SiO$_2$. Figure 2a shows the gate voltage dependence of the three components of the resistivity ($\rho_0$, $\rho_A$ and $\rho_B$) corresponding to scattering by impurities, graphene LA phonons, and RIP scattering by SiO$_2$ phonons, respectively, near room temperature (RT) for three different graphene samples ($T = 330$K, 308K & 306K for Samples 1, 2 & 3, respectively; Sample 1 and Sample 2 are the same samples shown in Figure 1, and Sample 3 is a lower mobility sample for which we have limited temperature dependence data.) The residual impurity resistivity $\rho_0(V_g)$ is estimated, with an error not greater than 1.5%, by taking $\rho(V_g, T)$ at low temperature ($T = $ 29K, 16K & 20K for Sample 1, 2 & 3, respectively). The graphene LA phonon resistivity $\rho_A(306$ K$) = 1.2 \times 10^{-3}$ $h/e^2$ is obtained from the global fit to Eqn. 1 for Samples 1 and 2. The RIP scattering resistivity $\rho_B(V_g, T \approx $ RT$)$ is obtained by subtracting $\rho_A(T)$ and $\rho_0(V_g)$ from $\rho(V_g, T)$ for each sample. Though $\rho_0(V_g)$ varies by a factor of 1.7X among the three samples, the temperature-dependent resistivities $\rho_B(V_g, T)$ are nearly equal except very close to the MCP (see Supplementary Note); this verifies that the temperature-dependent resistivity terms $\rho_A$ and $\rho_B$ arise from phonon scattering which is disorder-independent. The power-law behavior of the activated contribution $\rho_B(V_g, 306\text{ K}) \propto V_g^{-1.04}$ can also clearly be seen.

Figure 2b shows the corresponding room temperature mobility $\mu = 1/ne\rho = 1/c_g V_g \rho$ calculated for each resistivity contribution in Figure 2a as a function of gate voltage. If the properties of graphene were limited by the intrinsic LA phonon scattering as the dominant intrinsic source of resistivity, the room-temperature intrinsic resistivity of graphene would be 30 $\Omega$, independent of carrier density, and the mobility would diverge at low carrier density as $n^{-1}$. At a technologically-relevant carrier density $n = 10^{12}$ cm$^{-2}$ ($V_g = 14$ V), the intrinsic mobility would then be $2 \times 10^5$ cm$^2$/Vs, higher than any known semiconductor. If the only extrinsic limit to the mobility of graphene on SiO$_2$ were due to RIP scattering, graphene on SiO$_2$ would still have a room temperature mobility of $4 \times 10^4$ cm$^2$/Vs, which compares favorably to the best InAs and InSb FETs[25]. The dominance of RIP scattering over LA phonon scattering at room temperature poses an interesting tradeoff; high-$\kappa$ dielectrics may be used to reduce the scattering contribution from defects (i.e. $\rho_0$) due to increased screening of the impurity potential, but will *increase* scattering due to RIP[12].

Figure 3 shows the temperature dependence of the mobility of Sample 1 and Sample 2 at $n = 10^{12}$ cm$^{-2}$ ($V_g = 14$ V), as well as the limits due to scattering by LA phonons, polar optical phonons of the SiO$_2$ substrate, and impurities. As shown in Figure 3, even for the cleanest graphene devices fabricated to date, impurity scattering is the still the dominant factor limiting the mobility for $T < 400$ K. For comparison, the temperature-dependent mobility in Kish graphite and pyrolytic graphite from ref. [26] are also shown; these are the two materials commonly used as sources for exfoliated graphene on $SiO_2$. The significantly higher mobility at low temperature in Kish and pyrolytic graphites compared to graphene is a strong indication that the impurity scattering in graphene on $SiO_2$ is not due to point defects present in the parent material, but rather is likely caused by charged impurities in the $SiO_2$ substrate[14, 27]. It is important to note that the closeness of the room-temperature mobility values for graphene and bulk graphite is a coincidence, and removing impurity scattering in graphene will greatly increase not only the low temperature mobility, but the room temperature mobility as well.

Our data give a complete picture of the current limitations and future promise of graphene as an electronic material. Currently, mobility of graphene on $SiO_2$ at low and room temperature is limited by impurity scattering, likely due to charged impurities in the $SiO_2$ substrate[14, 27]. If charged impurity scattering can be reduced, the room-temperature mobility, limited by extrinsic RIP scattering due to $SiO_2$ phonons, could be improved to $4 \times 10^4$ cm$^2$/Vs, comparable to the best field-effect transistors[25]. With proper choice of substrate, or by suspending graphene, the intrinsic limit of mobility of $2 \times 10^5$ cm$^2$/Vs at room temperature could be realized. This would allow ballistic transport over micron lengths, opening the possibility of new electronic devices based on quantum transport operating at room temperature.

Acknowledgements: We acknowledge stimulating discussions with S. Das Sarma, E. Hwang, S. Adam, S. Fratini, and E. D. WIlliams. We also thank E. D. Williams for use of UHV facilities. This work has been supported by the U.S. Office of Naval Research grant no. N000140610882 (CJ, SX, MSF), NSF grant no. CCF-06-34321

(MSF), and the NSF-UMD-MRSEC grant no. DMR 05-20471 (JHC). MI is supported by the Intelligence Community Postdoctoral Fellowship program.

**Methods**

Graphene was obtained from Kish graphite by mechanical exfoliation[2] on 300nm SiO$_2$ over doped Si (back gate), with Au/Cr electrodes defined by electron-beam lithography. Raman spectroscopy confirmed that the samples are single layer graphene[14, 28] (see Supplementary Information for device pictures and Raman spectra). After fabrication, the devices were annealed in H$_2$/Ar at 300 °C for 1 hour to remove resist residues[14, 29].

In order to eliminate possible effects on the resistivity due to temperature-dependent concentrations of adsorbates on the graphene[14, 15], all measurements were performed in ultra-high vacuum (UHV). The devices were mounted on a liquid helium cooled cold finger in a UHV chamber, allowing temperature control between 16 K and 490 K. Following a vacuum chamber bakeout, each device was annealed in UHV at 490 K overnight to remove residual adsorbed gases. Experiments were carried out at pressure lower than $2 \times 10^{-9}$ torr at 490 K and $1 \times 10^{-10}$ torr below 300 K. The device temperature was tuned from 485 K to room temperature using a heater installed on the cold finger, and controlled liquid helium flow was used to tune the device temperature from 290 K to 16 K, with resistivity vs. gate voltage ρ($V_g$) curves taken at various temperature points. Warming experiments were also performed, where the device temperature was raised from 16 K to 243 K by controlling the helium flow. Heater operation was avoided at low temperature to prevent outgassing of the coldfinger. Transport properties of the samples between cooling and warming are very reproducible, showing no detectable effect of residual gas absorbed on the samples during the experiment; the exception is that small differences in cooling and warming data are occasionally observed very near the minimum conductivity point (MCP); see Supplementary Note.

Resistivity measurements were performed using a standard four-probe technique and error in determining the aspect ratio (and hence the absolute magnitude of the resistivity) is estimated to be 10%[14]. Resistivity vs. gate voltage $\sigma(V_g)$ curves are shifted by a constant threshold voltage $V_{th}$ in order to define $V_g = 0$ as the MCP. $V_{th}$ is small ($V_{th}$ = 0 V for Sample 1 and -3 V for Sample 2) and does not change with temperature for cleaned samples that are outgassed sufficiently in UHV. Sample 3 was prepared the same way as Sample 1 and Sample 2, and then multiple potassium deposition and removal cycles were carried out in UHV resulting in an increased density of immobile impurities and lowered mobility[14]. $V_{th}$ = -8.2 V for Sample 3.

**Figure Captions**

**Figure 1 Temperature-dependent resistivity of graphene on SiO$_2$. a,b,** Resistivity of two graphene samples as a function of temperature $\rho(V_g,T)$ for seven different gate voltages from 10 to 60 V on a linear scale. The left panel displays data from Sample 1 with two different experimental runs, and the right panel displays data from Sample 2. Short-dashed lines are fits to linear $T$-dependence (Eqn. 1) with a single slope of $4.0 \pm 0.5 \times 10^{-6}$ $he^{-2}K^{-1}$ from the global best fit to both samples and all $V_g$. **c,d,** Same data on a logarithmic scale. In **c,d**, solid lines are fits to Equation 2a, which includes acoustic phonon scattering in graphene and optical phonon scattering due to the SiO$_2$ substrate. Short-dashed lines are fits to Equation 2b, which includes the same acoustic phonon scattering term and a single Bose-Einstein distribution for the temperature activated resistivity. In addition to the low-temperature resistivity $\rho_0$, and linear term determined above, only two additional global parameters in Equation 2a and three global parameters in Equation 2b are used to fit the seven curves each for two devices.

**Figure 2 Performance limits of graphene on SiO$_2$ at room temperature. a**, Three components of the resistivity (the residual resistivity $\rho_0$, the graphene longitudinal acoustic phonon resistivity $\rho_A$, and the SiO$_2$ remote interfacial phonon scattering resistivity $\rho_B$) of three graphene samples near room temperature. Solid lines are the residual resistivities $\rho_0(V_g)$ for three graphene samples (Sample 1 in red, Sample 2 in blue, and Sample 3 in black). The dark red short-dashed line shows $\rho_A(T = 306$ K$)$ obtained from the global fit to Eqn. 1 to the data in Figs. 1**a-b**. The dark green short-dashed line shows $\rho_B(T = 306$ K$)$ obtained from the global fit to Eqn. 2a to the data in Figs. 1**c-d**.

The red squares and blue circles are $\rho_A(T = 306\,K)$ obtained from fits to individual curves in Figs. 1a-b for Samples 1 and 2 respectively. The experimental SiO$_2$ remote interfacial phonon scattering resistivity $\rho_B(V_g, T \approx RT)$ is obtained by subtracting $\rho_A(T)$ and $\rho_0(V_g)$ from $\rho(V_g, T)$ for each sample. The curves shown are $\rho_B(T = 330K)$ for Sample 1 (red dashed line), $\rho_B(T = 308K)$ for Sample 2 (blue dashed line), and $\rho_B(T = 306K)$ for Sample 3 (black dashed line). **b**, Gate-voltage-dependent mobility limits corresponding to the different sources of resistivity from **a**. Solid lines are the estimated mobility limits from the total electron-phonon interaction (i.e. $\rho_A + \rho_B$) as a function of gate voltage ($V_g$) for Sample 1 (red solid line) and Sample 2 (blue solid line) and Sample 3 (black solid line) near room temperature. Long-dashed lines are the extrinsic mobility limits from SiO$_2$ surface phonon scattering near room temperature for the three samples, calculated from the corresponding $\rho_B(V_g, T \approx 300K)$ curves in **a**. The dark red short-dashed line is the intrinsic mobility limit due to scattering by longitudinal acoustic phonons in graphene at 300K.

**Figure 3 Temperature dependence of the mobility in graphene and graphite.** The temperature-dependent mobilities of graphene Sample 1 (red squares) and Sample 2 (blue triangles) at $V_g = 14$ V ($n = 10^{12}\,cm^{-2}$) are compared with Kish graphite (solid black circles, data from Sugihara et al.[26]) and pyrolytic graphite (open black circles, data from Sugihara et al.[26]). The mobility limits in graphene determined in this work for scattering by LA phonons (dark red solid line), remote interfacial phonon scattering (dark green short-dashed line), and impurity scattering (red and blue dashed lines) are shown. Red

and blue solid lines show the expected net mobility for each sample, given by the sum of each individual contribution in inverse, according to Matthiessen's rule.

Figure 1

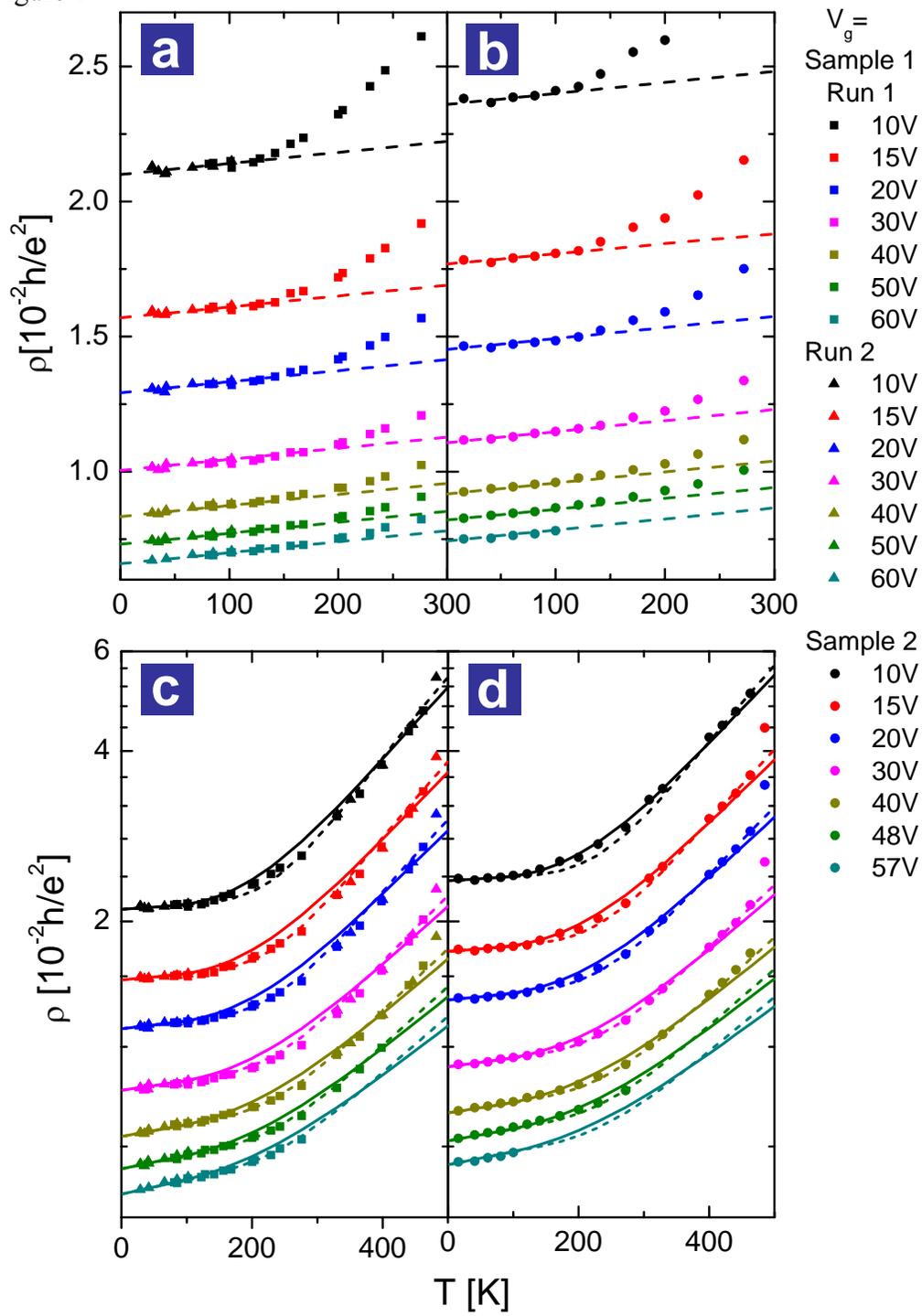

Figure 2

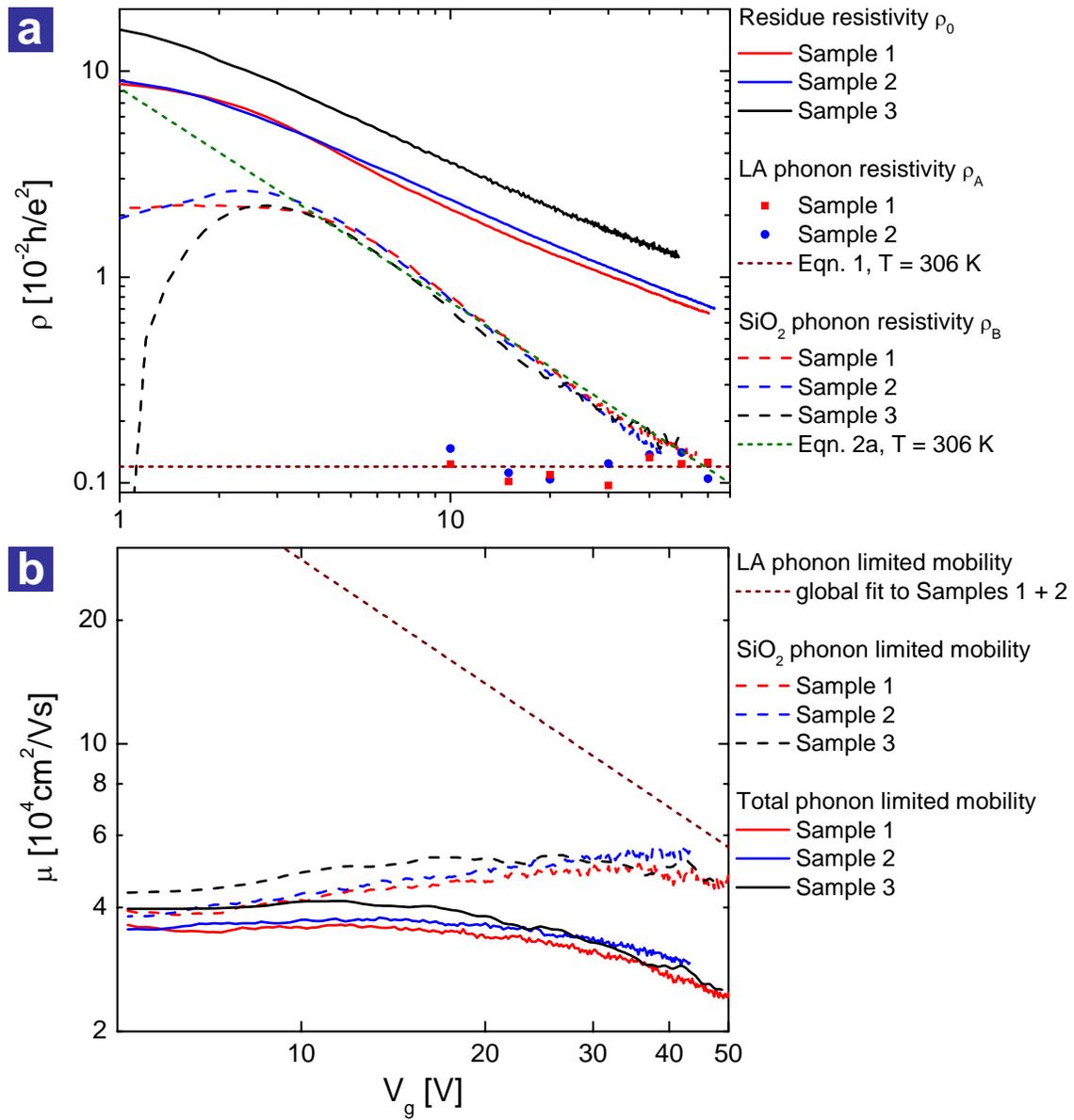

Figure 3

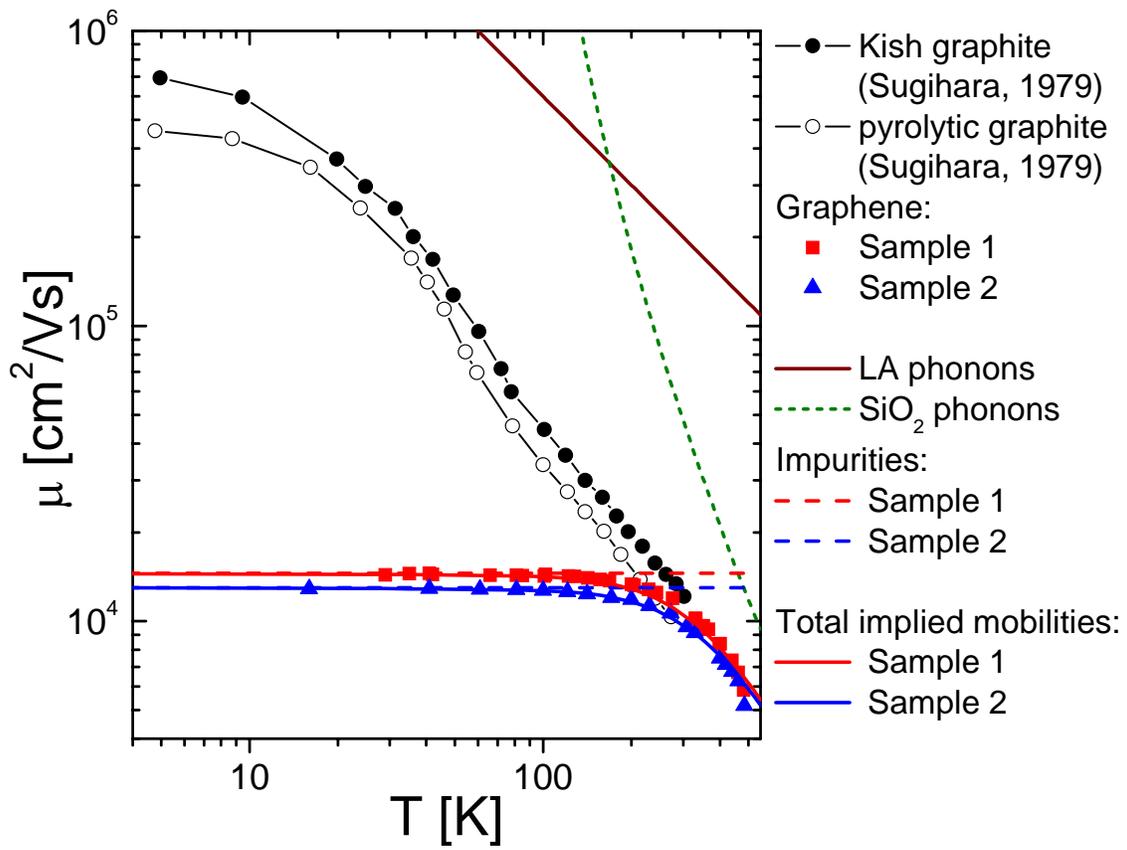

**Supplementary Notes**

**Sample Geometry and Raman Spectra**

Figures S1a - S1c show optical micrographs of the three devices used in this study. Figures S1d – S1f show the corresponding Raman spectra of the devices acquired over the device area using a confocal micro-Raman spectrometer with 633 nm excitation wavelength. The single Lorentzian 2D peak indicates the samples are single-layer graphene[1].

**Temperature Dependence of the maximum resistivity**

The resistivity at the minimum conductivity point (MCP) behaves very differently from the resistivity at higher carrier density ($V_g$ > 10 V). Figure S2 shows the maximum resistivity as a function of temperature $\rho_{max}(T)$ for the two samples presented in Figure 2 and one lower-mobility sample (Sample 3) for which we have more limited temperature-dependent data. $\rho_{max}(T)$ is highly sample-dependent, increasing with $T$ for Samples 1 and 2, and decreasing with $T$ for Sample 3. The latter behavior is expected for increased screening of the impurity potential by excited carriers[2,3] and the relative size of this effect should depend on the impurity density. This effect is expected to scale with $T/T_F$, and hence should be largest near the MCP. Furthermore, the effect is predicted to be small for $T < T_F = [363 \text{ K}] \times [V_g(\text{V})]^{1/2}$, which is well-satisfied except very near the MCP; which justifies the exclusion of screening in the analysis of the temperature dependence at non-zero $V_g$. The data for $\rho_{max}(T)$ for Sample 1 are also slightly different on warming and cooling, perhaps due to gases adsorbed on the sample at low $T$, consistent with $\rho_{max}(T)$ being highly dependent on the disorder in the sample. Taken together, the $\rho_{max}(T)$ data suggest an interplay of impurity screening and phonon scattering; more work will be need to disentangle these effects.

**Supplementary Figure Captions**

**Figure S1 Optical micrographs and Raman Spectra of the three graphene samples.**  **a**, **b**, **c**, Optical micrographs of Sample 1, 2 and 3, respectively.  **d**, **e**, **f**, Raman spectra of Sample 1, 2 and 3, respectively.  The blue dots are fits to Lorentzian lineshapes.

**Figure S2 Temperature dependence of the maximum resistivity.**  The maximum resistivities $\rho(V_g = 0, T)$ of graphene Sample 1, 2 and 3 are shown as a function of temperature.  Sample 1 and Sample 2 show increasing conductivity with temperature, though the functional form differs from Equation 2.  Sample 3 has lower mobility than Sample 1 and Sample 2, and shows a decreasing conductivity with increasing temperature.

**Supplementary References**

**Supplementary Figure S1**

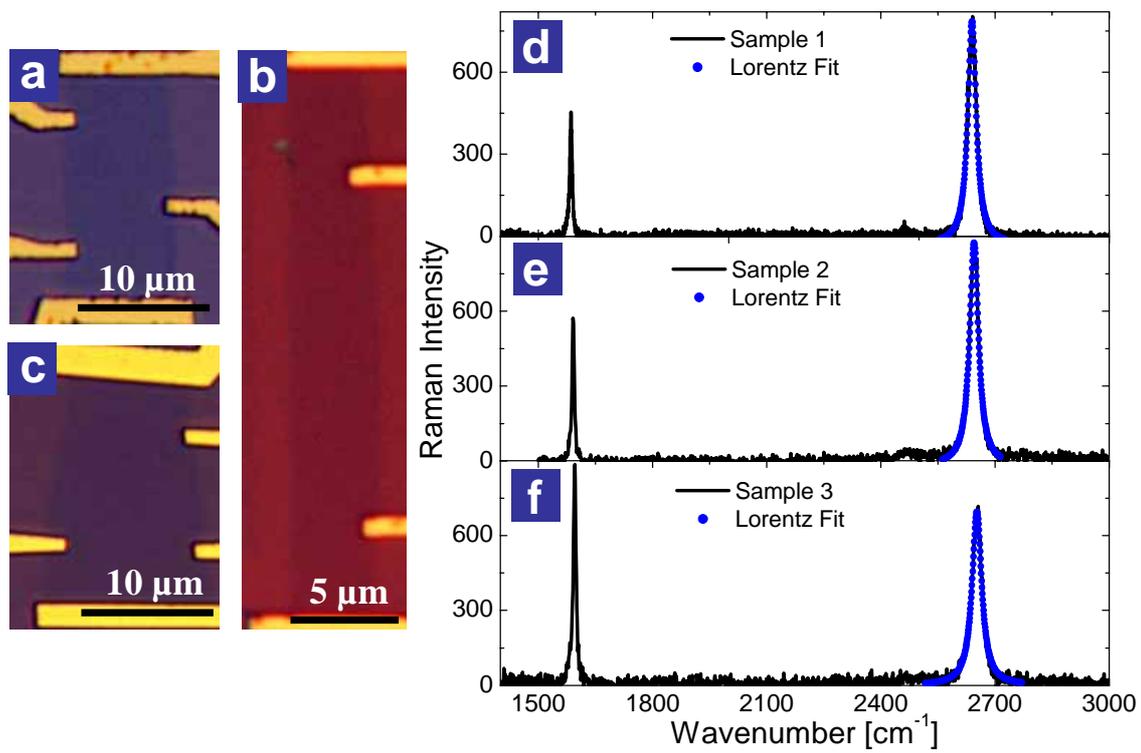

**Supplementary Figure S2**

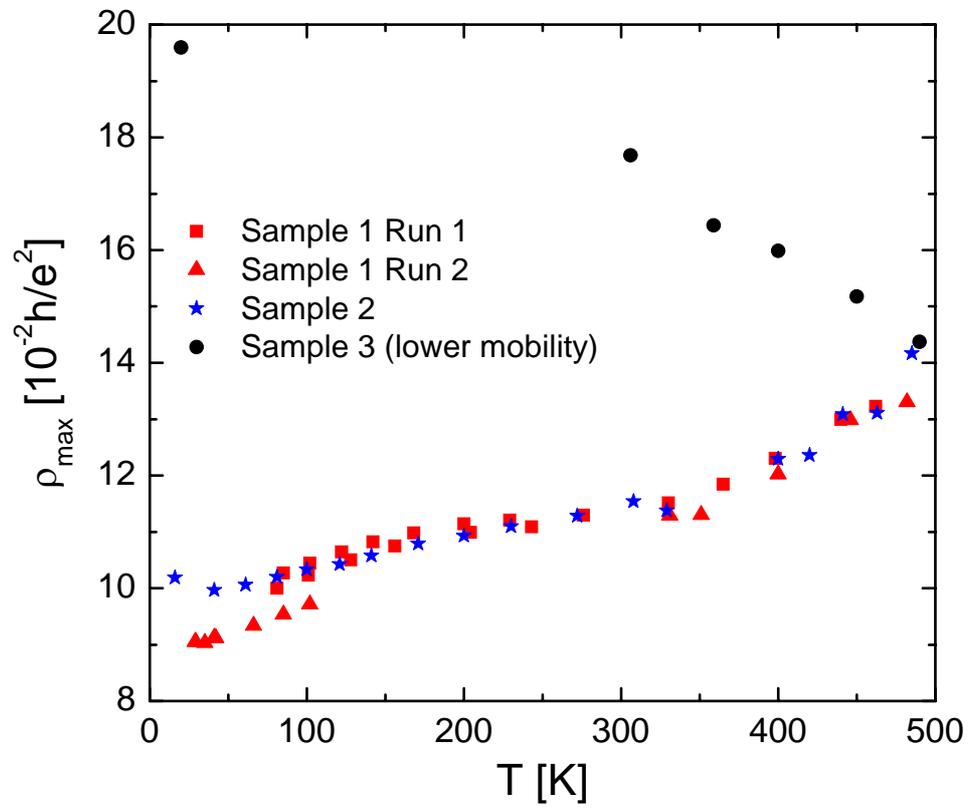